
\magnification=\magstep1
\looseness=2
\overfullrule=0pt
\hsize=6.5 truein
\vsize=9.0 truein
\baselineskip=12pt plus2pt
\parskip=6pt plus4pt minus2pt


\def\cl#1{\centerline{#1}}
\def\clb#1{\centerline{\bf #1}}
\def\bs{\bigskip}
\def\noi{\noindent}
\def\ltsima{$\; \buildrel < \over \sim \;$}
\def\lsim{\lower.5ex\hbox{\ltsima}}
\def\gtsima{$\; \buildrel > \over \sim \;$}
\def\gsim{\lower.5ex\hbox{\gtsima}}
\def\lcdm{$\Lambda$CDM}

\bigskip

\centerline{{\bf Whatever Happened to Hot Dark Matter?}
\footnote{*}{This article has been published in the SLAC magazine 
{\it Beam Line}, Fall 2001, Vol. 31, No. 3, pp. 50-57 (available
online at www.slac.stanford.edu\-/pubs/beamline/).  This is the text
as resubmitted, reflecting many improvements by the editor, Michael
Riordan, to whom I am grateful.}
}

\cl{by Joel R. Primack}
\cl{Physics Department, University of California, Santa Cruz, CA 95064 USA} 
\bs

The lightest of the fundamental matter particles, neutrinos may play
important roles in determining the structure of the universe.
Neutrinos helped to control the expansion rate of the universe during
the first few minutes, when the deuterium and most of the helium in
the universe were formed, and neutrinos and photons each accounted for
nearly half of the entire energy density of the universe during its
first few thousand years, before cold dark matter became gravitationally
dominant.  The fact that neutrinos are so ubiquitous --- with hundreds
of them occupying every thimbleful of space today --- means that they
can have an impact upon how matter is distributed in the universe.
Over the past two decades, the liklihood of this has risen and fallen
as more and more data has become available from laboratory experiments
and the latest telescopes.  We now know from the Super-Kamiokande
atmospheric neutrino data that neutrinos have mass.  Current estimates
are that the total neutrino mass could be comparable to that of the
visible stars in the universe, or perhaps even larger.

Dark matter made of light neutrinos, with masses of a few eV, is
called by cosmologists ``hot dark matter'' (HDM).  (See Sidebar 1
for a summary of the various dark matter types and of the
corresponding cosmological models.)  By 1979, cosmologists had become
convinced that most of the matter in the universe is completely
invisible.  This gravitationally dominant component of the universe
was named ``dark matter'' by the astronomer Fritz Zwicky, who first
described evidence for dark matter in the Coma cluster of galaxies in
1933: the galaxies were moving much too fast to be held together by
the gravity of the visible stars there.  This phenomenon was
subsequently seen in other galaxy clusters, but since the nature of
this dark matter was completely unknown it was often ignored.  During
the 1970s, it became clear that the motion of stars and gas in
galaxies, and of satellite galaxies around them, required that dark
matter must greatly outweigh ordinary matter in galaxies.  The data
gathered since then provides very strong evidence that most of the
mass in the universe is dark matter.

Here's the HDM story in a nutshell: For a few years in the late 1970s
and early 1980s, hot dark matter looked like the best bet dark matter
candidate.  Such HDM models of cosmological structure formation led to
a top-down formation scenario, in which superclusters of galaxies are
the first objects to form, with galaxies and clusters forming through
a process of fragmentation.  Such models were abandoned by the
mid-1980s when it was realized that if galaxies form sufficiently
early to agree with observations, their distribution would be much
more inhomogeneous than it is observed to be.  Since 1984, the
successful structure formation models have been those in which most of
the mass in the universe is in the form of cold dark matter (CDM).
But the HDM stock rose again a few years later, and for a while in the
mid-1990s it appeared that a mixture of mostly CDM with 20-30\% HDM
gave a better fit to the observations than either pure HDM or pure
CDM.  This ``cold + hot dark matter'' (CHDM) theory could explain data
on nearby galaxies and clusters only if the cosmological matter
density $\Omega_m$ is large, either unity (a ``critical density''
universe) or close to it.  And like all $\Omega_m=1$ theories, CHDM
predicted that clusters and galaxies would mostly form at low
redshift.  This turned out to disagree with observations, as clusters
were discovered at higher and higher redshifts.  Now increasing
observational evidence favors $\Lambda$CDM --- i.e., CDM with
$\Omega_m \approx 1/3$ and a cosmological constant $\Lambda$ or some
other form of ``dark energy'' making up $\Omega_\Lambda \approx 2/3$
so that $\Omega_{tot} = \Omega_m + \Omega_\Lambda =1$ as implied by
recent observations of the cosmic background radiation.  The question
now is how much room there is for HDM.  At present, cosmology provides
the best available upper limit on the neutrino masses.

It's already clear from this brief summary that to describe the
possible role of neutrinos as dark matter, I will have to say a few
words about how structures such as galaxies formed as the universe
expanded.  The expansion itself is assumed to be described by our
modern theory of gravity and spacetime, Einstein's theory of general
relativity (GR).  Is this a good assumption?  There are wonderfully
precise tests of GR on the scales of binary pulsars and the solar
system, and on much larger scales the masses of clusters of galaxies
measured by gravitational lensing agree with the masses of the same
clusters determined by the velocities of the galaxies and gas in them.
On still larger scales, the accuracy of standard gravity theory is
verified by the agreement between the observed flows of galaxies and
the motions predicted by their observed distribution.  The success of
cosmological theory is the best test of GR on really large scales.
For example, we now have three independent ways of estimating the age
of the universe (see Sidebar 2: The Age of the Universe), and their
agreement suggests that GR works on the largest scales.

In order for structure to form in the expanding universe, there must
either have been some small fluctuations in density in the initial
conditions, or else some mechanism to generate such fluctuations
afterward.  The only such fluctuation generation mechanisms that have
been investigated are ``cosmic defects'' such as cosmic strings, and
we now know that the pattern of fluctuations produced by such defects
is inconsistent with the fluctuations in temperature observed on
angular scales of a fraction of a degree in the cosmic microwave
background (CMB) radiation.  On the other hand, the sort of
fluctuations predicted by the simplest cosmic inflation models ---
adiabatic fluctuations, in which all components of matter and energy
fluctuate together --- are in excellent agreement with the latest CMB
results from the BOOMERANG and MAXIMA balloon flights and the DASI
instrument at the South Pole, announced at the American Physical
Society meeting in April 2001.

The evolution of adiabatic fluctuations is easy to understand if you
just remember that gravity is the ultimate capitalist principle: the
rich always get richer and the poor get poorer.  What I mean by this
is that, although the average density of the universe steadly decreases
due to its expansion, the regions that start out with a little higher
density than average expand a little slower than average and become
relatively more dense, while those with lower density expand a little
faster and become relatively less dense.  When any region has achieved
a density about twice that of an average region of its size, it stops
expanding and begins to collapse --- typically first in one direction,
forming a pancake-shaped structure, and then in the other two
directions.  

I can now explain the reason for the first hot dark matter boom about
1990.  Improving upper limits on CMB anisotropies were ruling out the
previously favored cosmological model with only ordinary matter.
There was also evidence from an experiment in a Moscow laboratory that
the electron neutrino mass is about 20-30 eV, which would correspond
to $\Omega_m = 1$ if the Hubble parameter were $h\approx0.5$, a value
that was compatible with the data available then (see Sidebar 3:
Neutrino Mass and Cosmological Density).  In a cosmology in which most
of the dark matter is light neutrinos, fluctuations on galaxy scales
are erased by ``free streaming'' of the ``hot'' (i.e., relativistic)
neutrinos in the early universe.

Since ``free streaming'' is the key property of HDM, it is worth
explaining this in a little more detail.  One year after the Big Bang,
a region about one light-year across contained the amount of matter
(both ordinary and dark matter) in a large galaxy like our own Milky
Way.  But the temperature then was about 100 million degrees, and
correspondingly each particle had a thermal energy of $10^4$ eV.  This
is much higher than the rest energy $m(\nu) c^2$ of light neutrinos,
which would therefore be moving at nearly the speed of light then.  As
a result of their rapid motion these neutrinos would spread out, amd
any fluctuations in the density of neutrinos on the mass scale of
galaxies would soon have become smoothed out.

The first scales to collapse in a HDM universe would correspond to the
mass inside the cosmic horizon when the temperature drops to a few eV
and the neutrinos become nonrelativistic.  This mass turns out to be
of the order of $10^{16}$ times the mass of our sun (or about $10^4$
times the mass of our galaxy, including its dark halo).  Evidence was
just becoming available from the first large-scale galaxy surveys that
the largest cosmic structures --- ``superclusters'' --- have masses of
approximately the same size.  This at first sight appeared to be a big
success for the HDM scenario.

Superclusters of roughly pancake shape were found observationally to
surround roughly spherical voids (regions where few galaxies are
found), in agreement with the first cosmological computer simulations,
which were run for the HDM model.  In the HDM model superclusters are
the first structures to form, since any smaller-scale fluctuations in
the dominant hot dark matter were erased by free-streaming.  Galaxies
must then form by fragmentation of the superclusters.  But it was
already clear that galaxies are much older than superclusters,
contrary to what the HDM scenario implies.  And the apparent detection
of electron neutrino mass by the Moscow experiment was soon
contradicted by results from other laboratories.

The cold dark matter model was developed in 1982-84 (partly by the
author of this article and his colleagues), just as the problems with
the hot dark matter model were becoming clear.  Proto-galaxies form
first in a CDM cosmology, and galaxies and larger-scale objects form
by aggregation of these smaller lumps --- although the cross-talk
between smaller and larger scales in the CDM theory naturally leads to
galaxies in clusters forming earlier than those in lower density
regions.  In this and other respects, CDM appeared to fit the
observational evidence much better than HDM.  The first great triumph
of CDM was that it successfully predicted (to within a normalization
uncertainty factor of about 2) the magnitude of the CMB temperature
fluctuations, which were discovered in 1992 using the COBE satellite.
But the simplest version of CDM, SCDM with $\Omega_m=1$, had already
begun to run into trouble.

Cosmological theories predict statistical properties of the universe
--- for example, the amplitude of fluctuations in the density of
matter on different scales, described mathematically by the power
spectrum.  Sound or other fluctuation phenomena can be described the
same way --- for example, low frequencies might be loud
(long-wavelength power).  The simplest way of describing the problem
with SCDM is to say that with a given amount of fluctuation power on
the large scales probed by COBE (billions of light years), it has a
little too much power on small scales relevant to the formation of
individual galaxies and clusters (millions of light years and less).
The fact that the SCDM theory could almost work across such a wide
range of size scales suggested that it had a kernel of truth.  The
question then was whether some variant of SCDM might work better.

I personally first became worried that SCDM was in trouble when the
large-scale flows of galaxies were first observed by my UCSC colleague
Sandra Faber and her ``Seven Samurai'' group of collaborators.  It had
earlier been established that the local group is moving at a velocity
of about 600 km/s with respect to the cosmic background radiation
reference frame.  But when the Seven Samurai and others found that
bulk motions of galaxies with similar velocities across regions
several tens of millions of light years across were the common
pattern, it became clear that this was inconsistent with the
expectations of the ``biased'' SCDM model that seemed to fit the
properties of galaxies on smaller scales.  So when the space shuttle
Challenger exploded at launch in January 1986 and as a result Hubble
Space Telescope could not be launched for several years, Jon Holtzman,
then Faber's graduate student, did a theoretical dissertation with me
instead of the HST-based observational dissertation he and Faber had
planned.  Holtzman's thesis, published in 1989, included detailed
predictions for 96 variants of CDM.  When we compared these
predictions with the data available in 1992, it was clear that the
best bets were $\Lambda$CDM and CHDM, each of which had less power on
small scales than SCDM.  Both of these variants had been proposed in
1984, when cold dark matter was still a new idea, but they were not
worked out in detail until a few years later when the problems with
SCDM began to surface.

Even if most of the dark matter is of the cold variety, a little hot
dark matter can have dramatic effects on the small scales relevant to
the formation and distribution of galaxies.  In the early universe,
the free streaming of fast-moving neutrinos washes out any
inhomogeneities in their spatial distribution on the scales that will
later become galaxies, just as in the HDM scenario.  As a consequence,
the growth rate of cold dark matter fluctuations is reduced on these
scales, and at the relatively late times when galaxies form there is
less flutuation power in CHDM models on small scales.  Since the main
problem with $\Omega_m=1$ cosmologies containing only cold dark matter
plus the usual ordinary matter (baryonic) contribution $\Omega_b \approx
0.04$ is that the amplitude of the galaxy-scale inhomogeneities is too
large compared to those on larger scales, the presence of a little hot
dark matter appeared to be possibly just what was needed.

And there was even a hint from an accelerator experiment that neutrino
mass might be in the relevant range.  The experiment was the Liquid
Scintillator Neutrino Detector (LSND) experiment at Los Alamos, which
saw a number of events that appear to be $\bar\nu_\mu \rightarrow
\bar\nu_e$ oscillations followed by $\bar \nu_e + p \to n + e^+$, 
$n + p \to D + \gamma$, with coincident detection of $e^+$ and the 2.2
MeV neutron-capture $\gamma$-ray.  Comparison of the LSND data with
exclusion plots from other experiments allows two discrete values of
$\Delta m^2_{\mu e}$, around 10.5 and 5.5 eV$^2$, or a range 2 eV$^2
\gsim \Delta m^2_{\mu e} \gsim 0.2$ eV$^2$.  The lower limit in turn
implies a lower limit $m(\nu) \gsim 0.5$ eV, or $\Omega_\nu \gsim
0.01$.  This would imply that the contribution of hot dark matter to
the cosmological density is greater than that of all the visible stars
$\Omega_\ast \approx 0.004$.  Such an important conclusion requires
independent confirmation.  The Karlsruhe Rutherford Medium Energy
Neutrino (KARMEN) experiment results exclude a significant portion of
the LSND parameter space, and the numbers quoted above take into
account the current KARMEN limits.  The Booster Neutrino Experiment
(BOONE) at Fermilab should attain greater sensitivity.

By 1995, supercomputer technology and simulation techniques had
advanced to the point where it was possible to do reasonably high
resolution cosmological-scale dissipationless simulations (i.e.,
without the hydrodyanamical complications of gas cooling, star
formation, and feedback from stars and supernovae) including the
random velocities of a hot dark matter component.  (The HDM
simulations five years earlier had actually been CDM simulations
starting from a HDM power spectrum.)  The results initially appeared
very favorable to CHDM [1].  Indeed, as late as 1998 a comprehensive
study of many CDM variants [2] found that a CHDM model with Hubble
parameter $h=0.5$ and density $\Omega_m=1$ including $\Omega_\nu=0.2$ was
the best fit to the galaxy distribution in the nearby universe of any
cosmological model.  But cosmological data was steadily improving, and
even by 1998 it had become clear that $h=0.5$ and $\Omega_m=1$ were
increasingly inconsistent with several sorts of observations, and that
instead $h \approx 0.7$ and $\Omega_m \approx 1/3$.  For example, it
was clear from the beginning that CHDM (and any other $\Omega_m=1$
model with a power spectrum consistent with the observed galaxy
distribution) predicts that galaxies and clusters form at relatively
low redshift, but around 1998 increasing numbers of galaxies were
discovered at redshifts beyond 3 and clusters began to be discovered
at redshifts of 1 and beyond.  And the fraction of baryons in
clusters, together with the reasonable assumption that this fraction
is representative of the universe as a whole, again gives $\Omega_m
\approx 1/3$.  That there is a large cosmological constant (or some
other form of dark energy) with $\Omega_\Lambda \approx 2/3$ then
follows from any two of the following three results: (1) $\Omega_m
\approx 1/3$, (2) the CMB anisotropy data implying that $\Omega_m +
\Omega_\Lambda =1$, and (3) the high-redshift supernova data implying
that $\Omega_\Lambda - \Omega_m \approx 0.4$.  

The abundance of galaxies and clusters at high redshifts is in
excellent agreement with the predictions of the $\Lambda$CDM model.
However, the highest resolution simulations of $\Lambda$CDM that were
possible in the mid-1990s gave a dark matter power spectrum that had
more power on scales of a few million light years than the observed
galaxy power spectrum on those scales, although they agreed on larger
scales [3].  This was inconsistent with the expectations that galaxies
would be if anything more clustered (or ``biased'') than the dark
matter on these scales, not less clustered.  However, when it became
possible to do still higher resolution simulations that allowed the
identification of the dark matter halos of individual galaxies, their
power spectrum turned out to be in excellent agreement with that of
the galaxies [4].  The galaxies were less clustered than the dark
matter because galaxies merged or were destroyed in very dense regions
because of interactions with each other and with the cluster center.
This explained why the galaxy power spectrum is so much lower than the
dark matter power spectrum on cluster scales, and it turned the former
disagreement into a triumph for $\Lambda$ CDM.

Thus $\Lambda$CDM is certainly the favored theory today.  But we know
from the atmospheric neutrino oscillations that there is enough
neutrino mass to correspond to some hot dark matter, at least
$\Omega_\nu \geq 10^{-3}$ (see Sidebar 3: Neutrino Mass and
Cosmological Density).  So the remaining question regarding neutrinos
in cosmology is how much more room there is for hot dark matter in
$\Lambda$CDM cosmologies.  The answer is, perhaps ten times that much,
but probably not 100 times.  The reason there is any upper limit at
all from cosmology is because the free-streaming of neutrinos in the
early universe slows the growth of the remaining cold dark matter
fluctuations on small scales, so to form the structure we see on the
scale of galaxies there must be much more cold than hot dark matter.
For the observationally favored range $0.2 \leq \Omega_m \leq 0.5$,
the latest comprehensive analysis [5] gives a limit on the sum of the
neutrino masses $m(\nu) \lsim 2.4(\Omega_m/0.17-1)$ eV (95\% C.L.), so
for $\Omega_m < 0.5$, $m(\nu) \lsim 5$ eV, and $\Omega_\nu \lsim 0.1$.
Astronomical observations that may soon lead to stronger upper limits
on $m(\nu)$ --- or perhaps a detection of neutrino mass --- include
data on the distribution of low-density clouds of hydrogen (the
``Lyman-alpha forest'') at high redshifts $z \sim 3$, large-scale weak
gravitational lensing data, and improved measurements of the cosmic
background radiation temperature fluctuations on small angular scales.
These types of data can be used to probe for the effects of any
free-streaming of neutrinos in the early universe which as we saw can
lead to less power on small scales, depending on the values of the
neutrino masses.  

The hot dark matter saga thus illustrates once again the fruitful
marriage between particle physics and cosmology: while neutrino
oscillation experiments can only tell us about the differences of the
squared masses of neutrinos, cosmology can tell us about the masses
themselves.  In an earlier example of this connection, cosmological
arguments based on Big Bang nucleosynthesis of light elements put a
strict limit on the possible number of light neutrino species; this
limit was eventually borne out high energy physics experiments on Z
bosons at CERN and SLAC.  The detailed studies of cosmological
structures now going on or about to begin may eveutually reveal
something about neutrino mass itself.

\vfil\eject

\clb{Sidebar 1: Summary of Dark Matter Types and Associated
Cosmological Models}


$$\vbox{ \settabs 3 \columns \+ {\it Dark Matter Type} & {\it Fraction of
critical density} & {\it Comment} \cr
\+ Baryonic DM &$\Omega_b \approx 0.04$        & about 10x visible matter \cr
\+ Hot DM      &$\Omega_\nu \approx 0.001-0.1$  & light neutrinos \cr
\+ Cold DM     &$\Omega_c \approx 0.3$         & most dark matter in
galaxy halos \cr}$$

$$\vbox{ \settabs 
\+ \lcdm\ \quad & CDM with $\Omega_{cdm} \approx 1/3$ and cosmological
constant $\Omega_\Lambda \approx 2/3$ \quad & 1984-2000 \cr
\+ {\it Acronym} & {\it Cosmological Theory} & {\it Flourished} \cr
\+ HDM & Hot DM cosmology with $\Omega_{tot}=1$ & 1978-1984 \cr
\+ SCDM & (standard) Cold DM with $\Omega_{tot}=1$ & 1982-1992 \cr
\+ CHDM & Cold + Hot DM with $\Omega_{cdm}\approx 0.7$ and $\Omega_\nu=0.2-0.3$ & 1994-1998 \cr
\+ \lcdm\ & CDM with $\Omega_{cdm} \approx 1/3$ and cosmological constant $\Omega_\Lambda \approx 2/3$ & 1984- \cr}$$

\bs\bs
\clb{Sidebar 2: The Age of the Universe}

In the mid-1990s there was a crisis in cosmology, because the age of
the old globular cluster (GC) stars in the Milky Way, then estimated
to be $t_{GC} = 16\pm3$ Gyr, was higher than the expansion age of the
universe, which for an $\Omega_m = 1$ universe is $t_{expansion}=
9\pm2$ Gyr.  Here I have assumed that the Hubble parameter has the
value $H_0$/(100 km/s/Mpc)$\equiv h=0.72\pm0.07$, the final result
from the Hubble Space Telescope project measuring $H_0$.

But when the data from the Hipparcos astrometric satellite became
available in 1997, it showed that the distance to the GCs had been
underestimated, which implied in turn that their ages had been
overestimated.  Correcting for this, and also using improved
treatments of stellar evolution, the age of the oldest GCs is
decreased to $t_{GC} = 13\pm3$ Gyr.  The age of the universe is then
$t_U \approx t_{GC} + \sim 1 {\rm Gyr \ for \ GC \ formation} 
\approx 14\pm3$ Gyr.

Several lines of evidence now show that the universe does not have
$\Omega_m = 1$ but rather $\Omega_m + \Omega_\Lambda = 1.0\pm0.1$ with
$\Omega_m = 0.3\pm0.1$.  Lowering $\Omega_m$ increases the expansiion
age, and a cosmological constant $\Omega_\Lambda > 0$ increases it
still further, so now $t_{expansion} = 13\pm2$ Gyr.  This is now in
excellent agreement with the globular cluster age.  The high-redshift
supernova data alone give an expansion age $t_{SN} = 14.2\pm1.7$ Gyr.

Moreover, a new type of age measurement based on radioactive decay of
Thorium-232 (half-life 14.1 Gyr) measured in a number of very old
stars gives a completely independent age $t_{Th} = 14\pm3$ Gyr. A
similar measurement, based on the first detection in a star of
Uranium-238 (half-life 4.47 Gyr), is reported in the 12 Feb 2001 issue
of Nature, giving $t_U = 12.5\pm3$ Gyr.  Work in progress should soon
improve the precision of this sort of measurement.

All the recent measurements of the age of the universe are thus in
excellent agreement.  It is reassuring that three completely different
clocks --- stellar evolution, expansion of the universe, and
radioactive decay --- agree so well.

\bs\bs
\clb{Sidebar 3: Neutrino Mass and Cosmological Density}

Cosmic rays hitting the top of the atmosphere all around the world
produce ``atmospheric neutrinos.''  The atmospheric neutrino data from
the Super-Kamiokande underground neutrino detector in Japan provide
strong evidence of muon to tau neutrino oscillations, and therefore
that these neutrinos have nonzero mass.  This is now being confirmed
by the K2K experiment, in which a muon neutrino beam from the KEK
accelerator is directed toward Super-Kamiokande and the number of
muon neutrinos detected is just as expected with the muon neutrinos
oscillating to tau neutrinos at the atmospheric rate.  

Oscillation experiments cannot measure neutrino masses directly, only
the squared mass difference $\Delta m^2_{ij} \equiv |m_i^2 - m_j^2|$
between the oscillating species.  The Super-Kamiokande atmospheric
neutrino data imply that $ 5 \times 10^{-4} < \Delta m^2_{\tau \mu} <
6 \times 10^{-3}$ eV$^2$ (90\% CL), with a central value $\Delta
m^2_{\tau \mu} = 3 \times 10^{-3}$ eV$^2$.  If the neutrinos have a
hierarchical mass pattern $m(\nu_e) \ll m(\nu_\mu) \ll m(\nu_\tau)$
like the quarks and charged leptons, then this implies that $\Delta
m^2_{\tau \mu} \approx m(\nu_\tau)^2$, so $m(\nu_\mu) \approx 5 \times
10^{-2}$ eV.  These data imply a lower limit on the HDM (i.e., light
neutrino) contribution to the cosmological density $\Omega_\nu \gsim
0.001$ --- almost as much as that of all the stars in the disks of
galaxies --- and permit higher $\Omega_\nu$.

There is a connection between neutrino mass and the corresponding
contribution to the cosmological density, because thermodynamics in
the early universe determines the abundance of each of the three
neutrino species (including both neutrinos and antineutrinos) to be
about 112 per cubic centimeter.  It follows that the density
$\Omega_\nu \equiv \rho_\nu / \rho_c$ contributed by neutrinos, in
units of critical density $\rho_c = 10.54 \, h^2 {\rm keV}\,{\rm
cm}^{-3}$, is $\Omega_\nu = m(\nu) /( 93\, h^2 {\rm eV})$, where
$m(\nu)$ is the sum of the masses of all three neutrino species.
Since $h^2 \approx 0.5$, $m(\nu_\tau) \approx 0.05$ eV corresponds to
$\Omega_\nu \approx 10^{-3}$.

However, this is a lower limit, since in the opposite case where the
oscillating species have nearly equal masses, the values of the masses
themselves could be much larger.  The only other laboratory approaches
to measuring neutrino mass are (1) the attempt to detect neutrinoless
double beta decay, which is sensitive to the value of a possible
Majorana component of the electron neutrino mass, and (2) precise
measurements of the endpoint of the tritium beta decay spectrum, which
give an upper limit on the mass of the electron neutrino, given in the
online 2001 Particle Data Book as 3 eV.  Because of the small values
of both squared mass differences $\Delta m^2_{e\mu} \lsim 10^{-5}$
eV$^2$ from solar neutrino oscillations and $\Delta m^2_{\mu \tau}
\lsim 6 \times 10^{-4}$ eV$^2$ from atmospheric neutrino oscillations
as discussed above, the tritium upper limit on $m(\nu_e)$ becomes an
upper limit on all three neutrino species.  But this is not a very
stringent upper limit, corresponding to a maximum total neutrino mass
$m(\nu) < 9$ eV.  Perhaps surprisingly, cosmology already provides a
stronger constraint on $m(\nu) \lsim 5$ eV, from the effects of
neutrinos on structure formation discussed in this article.

\vfill\eject
\clb{References}

\noi
[1] Joel R. Primack, Jon Holtzman, Anatoly Klypin, and
David O. Caldwell, {\sl Phys. Rev. Lett.} {\bf 74}, 2160 (1995).

\noi
[2] Eric Gawiser and Joe Silk, {\sl Science} {\bf 280}, 1405 (1998).

\noi
[3] Anatoly A. Klypin, Joel R. Primack, and Jon Holtzman, {\sl
Astrophys. J.} {\bf 466}, 13 (1996); A. Jenkins et al., {\sl
Astrophys. J.} {\bf 499}, 20 (1998).

\noi
[4] Anatoly A. Klypin et al., {\sl Astrophys. J.} {\bf 516}, 530
(1999); Pedro Colin et al.,  {\sl Astrophys. J.} {\bf 523}, 32 (1999).

\noi
[5] Rupert A.C. Croft, Wayne Hu, and Romeel Dav\'e, {\sl
Phys. Rev. Lett.} {\bf 83}, 1092 (1999).  Cf. X. Wang, M. Tegmark, and
M. Zaldarriaga, {\sl Phys. Rev. D} in press, astro-ph/0105091.

\bigskip
\leftline{\bf Additional References}

\noi
A technical article on the present subject with many references:
``Hot Dark Matter in Cosmology,'' by Joel R. Primack \& Michael
A. K. Gross, in {\it Current Aspects of Neutrino Physics},
ed. D. O. Caldwell (Berlin: Springer, 2001) pp. 287-308; also available
on the web as astro-ph/0007165 and interactively as
nedwww.ipac.caltech.edu/level5/Primack4/frames.html

\noi
The Neutrino Oscillation Industry webpage provides convenient links to
current and future neutrino experiments:
www.hep.anl.gov/ndk/hypertext/nuindustry.html

\noi
For reviews of the current status of cosmology see, e.g.,
``Cosmological Parameters 2000,'' by Joel R. Primack, in {\it Sources
and Detection of Dark Matter in the Universe}, Proc. 4th International
Symposium (DM 2000), Marina del Rey, California, 20-23 Feb 2000,
ed. D. Cline (Berlin: Springer, 2001), pp. 3-17, astro-ph/0007187; and
``The Nature of Dark Matter,'' by Joel R. Primack, to appear in {\it
Proceedings of International School of Space Science 2001}, ed.  Aldo
Morselli (Frascati Physics Series), astro-ph/0112255.

\bye